\documentclass[12pt]{iopart}
\eqnobysec
\begin{document}

\title[Equations of state for solids from shock-wave data]{Equations
of state for solids at high pressures and temperatures
from shock-wave data}

\author{Valentin Gospodinov\footnote{\it e-mail:
 valentin@pronto.phys.bas.bg}}

\address{Bulgarian Academy of Sciences, Space Research Institute,
Space Materials Division,
1000 Sofia  P.O. Box 799, BULGARIA}

\begin{abstract}
     This paper deals with the analytic derivation of a complete equation of
state (EOS) for solids from shock-wave data in the range of pressures and
temperatures, attained by detonation of chemical explosives. It is assumed
that stresses behind the shock wave are isotropic, and the compressed
materials behind the shock front are in thermodynamic equilibrium. A single
Hugoniot curve determined from shock-wave experiments does not provide enough
thermodynamic information to specify an EOS. The assumptions which, along
with the shock-wave data, are used in the present work to determine a
complete EOS are the Gr\"uneisen assumption, the volume dependence of the
Gr\"uneisen parameter, and the Debye model for the specific heat.
Consequently, the thermodynamic functions of the system under consideration
are represented as a superposition of a cold (0~{\em K}\/) term corresponding
to the cold compressed state and a second one corresponding to the thermal
vibrations of atoms in the crystal lattice about their mean positions. A
differential equation is derived for the cold isotherm of the internal
energy, using the Mie-Gr\"uneisen model and the shock Hugoniot as a reference
curve. An analytic solution to the equation is obtained. Based on this
solution, the complete, the caloric, the thermal and the incomplete EOS are
determined. Also, an equation is derived for the shock temperature along
the Hugoniot curve.
\end{abstract}


\date{today}
\maketitle

\section*{Introduction}

The study of shock-wave propagation in solids has added significantly to our
understanding of physical processes which take place at high pressures, high
temperatures and very short times. One particular aspect of these studies is
to determine the {\em equations of state (EOS)} for different materials from
shock-wave data. Shock-wave experiments, aimed at EOS investigations, render
it possible to extend the range of pressure-volume data beyond the region
that can be reached with conventional static pressure experiments and to
use much bigger specimens of the investigated material.

Equations of state of matter, obtained both theoretically and experimentally
are of immense current importance in the basic and the applied sciences.
They describe the dependence of thermodynamic properties of matter on its
microscopic internal structure and have a variety of applications. These
include studies of the state of matter in the Earth's and planetary
interiors, shock-induced chemical reactions and polymorphic phase
transformations, hypervelocity impact, powder compaction or other
applications as well as their computer simulation based on hydrodynamic
models. In principle, all the thermodynamic properties of a system can be
predicted once precise knowledge of the complete EOS becomes known. But a
single Hugoniot curve determined from shock-wave experiments does not
provide enough thermodynamic information to specify an EOS. The assumptions
which, along with shock-wave data, are used in this work to determine a
complete EOS are the Gr\"uneisen assumption (1), the volume dependence of the
Gr\"uneisen parameter (2), and the Debye model for the specific heat (3).
Also, it is assumed that stresses behind the shock wave are isotropic,
and the compressed materials behind the shock front are in thermodynamic
equilibrium.

Therefore the objective of the work presented herein is to obtain a complete
EOS for solids from shock-wave data using the Debye - Mie-Gr\"uneisen model
and the shock Hugoniot as a reference curve, and to establish the
thermodynamic properties of a system with such an EOS. Also, an equation for
the shock temperature along the Hugoniot curve will be derived.

\section{General formulation}

Formally, the equation of state is a functional relationship among the
thermodynamic variables for a system in equilibrium. One of the most
frequently used thermodynamic relations is the {\em thermal EOS}, or just
the EOS. For a simple thermodynamic system it has the form
\begin{equation}  \label{one}
f(P,V,T)=0.
\end{equation}

\noindent It represents a surface in the {\em PVT} space. The {\em PV}
isotherm ({\em T} = {\em const.}), isentrope ({\em S} = {\em const.}) and
the shock Hugoniot (the locus of all possible states that can be reached by
using a single shock from a given initial state) are particular curves on
this surface.

To obtain the form of \ {\em f}, it is convenient to derive the Helmholtz
free energy of the system
\begin{equation}  \label{two}
F=E-TS
\end{equation}

\noindent and compute {\em P} as the volume derivative
\begin{equation}  \label{three}
P=-(\partial F/\partial V)_{T.}
\end{equation}

It is clear that {\em F} depends upon the microscopic structure of the solid
under consideration, which would vary as a function of volume and
temperature. At different temperatures and densities, the corresponding
region of matter will be dominated by different interactions.

In view of the above, we may write the free energy as a superposition of
terms appropriate to various physical phenomena (Gr\"uneisen's assumption
\cite{Gruneisen})
\begin{equation}  \label{four}
F(V,T)=F_c(V)+F_{vib}(V,T)+F_e(V,T)
\end{equation}

\noindent Here {\it F}$_c$ is the configurational free energy at 0{\it K}.
$F_{vib}$ corresponds to the contribution of the zero and thermal vibrations
of the ions of the crystal lattice. $F_e$ is the conduction electron
thermal excitations contribution.

Expressions for the internal energy, pressure and other thermodynamic
functions are similar to Eq.(\ref{four}):
\begin{equation}
P=P_c+P_{vib}+P_e,\ E=E_c+E_{vib}+E_e.  \label{five}
\end{equation}

\noindent The contribution of the conduction electrons thermal excitations
manifests itself at temperatures $\geq $ 10$^4$ {\em K} and pressures $\geq $
10$^2$ {\em GPa} \cite{Godwal}. At lower temperatures and pressures it may
be neglected and Eqs.(\ref{four}) and (\ref{five}) take the form
\begin{equation}
F=F_c+F_{vib},\ P=P_c+P_{vib},\ E=E_c+E_{vib}.  \label{six}
\end{equation}

\noindent The vibrational terms are sums of the quantum zero-degree and the
thermal vibrations of the crystal lattice. Then Eqs.(\ref{six}) may be
written as
\begin{equation}
\begin{array}{c}
F(V,T)=F_c(V)+F_0(V)+F_{lT}(V,T), \\
E(V,T)=E_c(V)+E_0(V)+E_{lT}(V,T), \\
P(V,T)=P_c(V)+P_0(V)+P_{lT}(V,T).
\end{array}
\label{seven}
\end{equation}

\noindent {\em F}$_0$, {\em E}$_0$ and {\em P}$_0$ are the quantum
zero-degree contributions. They are functions of volume only, like the
configurational terms. {\em F}$_{lT}$, {\em E}$_{lT}$ and {\em P}$_{lT}$ are
the lattice thermal contributions. They depend on volume and temperature.

The pressures and temperatures attained by detonation of chemical explosives
in contact with the specimen or explosively accelerated flyer plate methods
do not exceed 10$^4$ {\em K} and 10$^2$ {\em GPa}. These techniques are
the most accessible to the general scientific community and because of
that the most frequently used ones.
The behaviour of a solid within this range of pressures and
temperatures is determined by the lattice thermal vibrations. That is
why it is often called the \textsl{phonon region}. It looks narrow on
the background of the highest temperatures and pressures, investigated
by contemporary physics. But in addition to being easily accessible, it
covers a wide variety of processes and phenomena of interest to science
and technology. Those mentioned in the introduction are to name but
few. Consequently, we shall aim our considerations at the phonon region
and seek the explicit form  of Eqs.(\ref{seven}).

From Eq.(\ref{two}) and the fundamental thermodynamic identity
\[
dE=-PdV+TdS
\]

\noindent it follows
\begin{equation}  \label{eight}
E_c=F_c\quad \ \mbox{\rm and}\ \quad P_c=-dF_c/dV=-dE_c/dV.
\end{equation}

\noindent The specific form of Eqs.(\ref{eight}) will be derived from the
Mie-Gr\"{u}neisen equation using the shock Hugoniot as a reference curve.

The calculation of Helmholtz free energy components {\em F}$_0${\em \ (V)}
and {\em F}$_{lT}${\em \ (V,T)} will be performed within the framework of
the Debye model for the specific heats.

\section{Calculation of the zero-degree Kelvin isotherm from shock-wave data}

The shock-wave methods for deriving the cold isotherm are based on the
measurement of the Hugoniot curve. The quantities directly measured are the
kinematic parameters of the shock wave - the shock front velocity {\em u}$_s$
and the particle velocity in the compressed region {\em u}$_p$ . The
relation {\em u}$_s${\em (u}$_p${\em )} is the shock Hugoniot. Most
substances in the absence of phase transitions have a linear shock Hugoniot
\cite{McQueen}
\begin{equation}  \label{nine}
u_s=c_0+su_p
\end{equation}

\noindent in a wide range of particle velocities. The Hugoniot intercept,
{\em c}$_0$ , and the slope, {\em s}, are determined from the data by the
method of least squares. If rigidity effects and possible low pressure phase
changes are neglected, the intercept should correspond to the velocity of an
infinitesimal pressure pulse, or the bulk sound speed, {\em c}$_0$ =$\left[
(\partial P/\partial \rho )_S\right] ^{1/2}$ (at {\em P} = 0). Since the
slope is linearly related to the pressure derivative of the adiabatic bulk
modulus, ($\partial ${\em B}/$\partial ${\em P})$_S$, a linear {\em u}$_s$%
{\em -u}$_p$ Hugoniot then reflects a nearly linear dependence of {\em B}$_S$
on the pressure.

The transition from kinematic {\em (u}$_s${\em ,u}$_p${\em )} to
thermodynamic {\em (P,V,E)} variables is done using the laws of conservation
of mass, momentum and energy across the shock front \cite{Zharkov}
\begin{equation}  \label{ten}
\rho _0u_s=\rho (u_s-u_p),
\end{equation}
\begin{equation}  \label{eleven}
P-P_0=\rho _0u_su_p,
\end{equation}
\begin{equation}  \label{twelve}
\left[ \left( E-E_0\right) -\frac{u_p^2}2\right] \rho _0u_s=P_0u_p.
\end{equation}

\noindent Here {\em E}, {\em P} and $\rho $ are the specific internal
energy, the pressure and the density behind the shock front, and {\em E}$_0$%
, {\em P}$_0$, $\rho _0$ are the values of these quantities ahead of the
shock front. From Eqs.(\ref{ten}) and (\ref{eleven}) we obtain
\begin{equation}
u_s=V_0\left[ (P-P_0)/(V_0-V)\right] ^{1/2},  \label{thirteen}
\end{equation}
\begin{equation}
u_p=\left[ (P-P_0)(V_0-V)\right] ^{1/2}.  \label{forteen}
\end{equation}

\noindent Solving Eqs.(\ref{thirteen}) and (\ref{forteen}) for {\em P} gives
\begin{equation}
P=P_0+\frac 1{V_0}u_su_p\ ,\ V=V_0\left( \frac{u_s-u_p}{u_s}\right) .
\label{fifteen}
\end{equation}

\noindent Equations (\ref{fifteen}) give the relationship between the
kinematic variables {\em u}$_s$ , {\em u}$_p$ and the thermodynamic variables
{\em P} and {\em V}.

Substituting with the right-hand sides of Eqs.(\ref{thirteen})-(\ref{forteen}%
) in Eq.(\ref{twelve}) gives the Hugoniot equation of energy
\begin{equation}
E-E_0=(P+P_0)(V_0-V)/2.  \label{sixteen}
\end{equation}

\noindent that defines all states on the ({\em E-P-V}) surface that can be
reached from an initial state ({\em E}$_0$, {\em P}$_0$,{\em V}$_0$) by a
single shock.

If the linear {\em u}$_s${\em -u}$_p$ relation holds, the Rankin-Hugoniot
equations (\ref{fifteen}) and (\ref{sixteen}) can be used to express
pressure and energy as functions of volume along the Hugoniot by the
following convenient analytic expressions (with {\em P}$_0$ and {\em E}$_0$
taken to be zero at ambient conditions)
\begin{equation}
P_H=\frac{\rho _0c_0^2\varepsilon }{\left( 1-\varepsilon s\right) ^2},
\label{seventeen}
\end{equation}
\begin{equation}
E_H=\frac{c_0^2\varepsilon ^2}{2\left( 1-\varepsilon s\right) ^2},
\label{eighteen}
\end{equation}
\begin{equation}
\varepsilon =1-V/V_0=1-\rho _0/\rho .  \label{nineteen}
\end{equation}

Equations (\ref{seventeen})-(\ref{eighteen}) and the values of {\em c}$_0$
and {\em s} summarize all the experimental thermodynamic information
which is available from shock-wave measurements.

To compute the cold compression curve we shall employ the Mie-Gr\"uneisen
equation \cite{McQueen} in the form
\begin{equation}  \label{twenty}
P(V,T)-P(V,0)=\frac{\gamma (V)}V\left[ E(V,T)-E(V,0)\right] ,
\end{equation}

\noindent where $\gamma ${\em (V)} is the Gr\"{u}neisen parameter. Equations
(\ref{seventeen},\ref{eighteen}) give the pressure {\em P}$_H$ and the
specific internal energy {\em E}$_H$ on the shock Hugoniot. Since it
connects equilibrium thermodynamic states we can write
\begin{equation}
P_H(V)-P(V,0)=\frac{\gamma (V)}V\left[ E_H(V)-E(V,0)\right] .
\label{twentyone}
\end{equation}

We shall assume that $\gamma $ is a function only of volume and,
specifically, that the product $\rho \gamma $ is constant. Experimental
work on a number of materials \cite{Carter}, as well as theoretical
considerations \cite{Walsh}, indicate this to be an adequate approximation.
With this assumption, $\gamma ${\em (V)} is given by
\begin{equation}  \label{twentytwo}
\gamma (V)/V=\gamma _0/V_0,
\end{equation}

\noindent where $\gamma _0$ is the thermodynamic value at standard
conditions given by $\gamma _0=3\alpha c_0^2/C_p$, $\alpha $ is the thermal
expansion coefficient, {\em c}$_0$ the sound speed, and {\em C}$_p$ the
specific heat at constant pressure.

At {\em T} = 0{\em K} {\em F = E} and from Eqs.(\ref{eight}) it follows
\begin{equation}
P_c(V)=P(V,0)=-dE_c(V)/dV.  \label{twentythree}
\end{equation}

If we substitute with Eqs.(\ref{seventeen}),(\ref{eighteen}),(\ref
{twentythree}) in the left-hand side of Eq.(\ref{twentyone}), we obtain a
differential equation for the zero Kelvin isotherm of the specific internal
energy {\em E}$_c${\em \ (}$\varepsilon ${\em ) = E(}$\varepsilon ${\em ,0)}:
\begin{equation}
\frac{dE_c(\varepsilon )}{d\varepsilon }-\gamma _0E_c(\varepsilon )-\frac{%
c_0^2\varepsilon (1-\gamma _0/2)}{(1-s\varepsilon )^2}=0,  \label{twentyfour}
\end{equation}
\[
E_c(\varepsilon =0)=E_{00}.
\]

\noindent This is a first order linear inhomogeneous ordinary differential
equation. The functional form of {\em E}$_c${\em \ (}$\varepsilon ${\em )}
may be obtained by integrating Eq.(\ref{twentyfour}). Unfortunately, the
solution cannot be given by combinations of elementary functions. It is
possible, however, that it can be approximated by a power series in $%
\varepsilon $ within high precision. In this case the cold compression curve
{\em E}$_c${\em (}$\varepsilon ${\em )} can be written as the following
power series in $\varepsilon $%
\begin{equation}
E_c(\varepsilon )=\sum\limits_iE_{0i}\varepsilon ^i,  \label{twentyfive}
\end{equation}

\noindent where $\varepsilon $ is the dimensionless volume given by Eq.(\ref
{nineteen}). {\em E}$_{00}$ is obtained from the assumed reference state and
the remaining coefficients to the fifth order in $\varepsilon $ are given by
the expressions:
\[
E_{01}=\gamma _0E_{00},\ E_{02}=\frac 12\left( c_0^2+\gamma
_0^2E_{00}\right) ,\ E_{03}=\frac 16\left( 4sc_0^2+\gamma _0^3E_{00}\right)
,
\]
\begin{equation}
E_{04}=-\frac 1{24}\left[ 2\left( \gamma _0-9s\right) c_0^2s-\gamma
_0^4E_{00}\right] ,  \label{twentysix}
\end{equation}
\[
E_{05}=-\frac 1{120}\left[ 2\left( \gamma _0^2+9\gamma _0s-48s^2\right)
c_0^2s-\gamma _0^5E_{00}\right]
\]

\noindent From Eqs.(\ref{twentythree}) and (\ref{twentyfive}) we can derive
an expression for the zero Kelvin isotherm of the pressure:
\begin{equation}
P_c(\varepsilon )=\sum\limits_iP_{0i}\varepsilon ^i.  \label{twentyseven}
\end{equation}

\noindent The coefficients {\em P}$_{0i}$ and {\em E}$_{0i}$ are not
independent. A relation exists between them which is given by
\begin{equation}  \label{twentyeight}
P_{0i}=(i+1)E_{0i+1}/V_0.
\end{equation}

\section{Contribution of the quantum zero Kelvin and lattice thermal
vibrations to the EOS}

In accordance with the quasiharmonic approximation the vibrational energy
levels of a crystal lattice with {\em N} atoms may be treated as the energy
levels of a system of 3{\em N} independent linear harmonic oscillators. For
the logarithm of the partition function of such a system we can write
\begin{equation}  \label{twentynine}
\ln Z=\sum_{i=1}^{3N}\ln z_i=\sum_{i=1}^{3N}\ln \frac{\exp \left( -h\nu
_i/2kT\right) }{1-\exp \left( -h\nu _i/kT\right) },
\end{equation}

\noindent where {\em z}$_i$ is the partition function of the {\em i}-th
oscillator.

In the Debye model \cite{Morse} the crystal lattice is replaced by an
isotropic elastic medium with a continuous dispersion law function. Further
it is assumed that the magnitude of frequency in such a system does not
exceed a certain boundary value, the Debye frequency $\nu _D$ , chosen so
that the number of the independent lattice oscillations is equal to the
total number 3{\em N} of the lattice degrees of freedom. It follows from
this approximation that we can replace the summation in Eq.(\ref{twentynine})
by integration. Let us omit the index {\em i} of the oscillations. Then
\[
\ln \frac{\exp \left( -h\nu /2kT\right) }{1-\exp \left( -h\nu /kT\right) }
\]

\noindent is ln{\em z} for all oscillators in the frequency range between $%
\nu $ and $\nu $+{\em d}$\nu $. The number of these oscillators is equal to
{\em g(}$\nu ${\em )d}$\nu $, where
\begin{equation}  \label{thirty}
g(\nu )=9N\nu ^2/\nu _D^3
\end{equation}

\noindent is the density of distribution of oscillators among frequencies
\cite{Morse}. We introduce the Debye temperature $\theta _D=h\nu _D/k$,
which depends on volume and is specific for each substance, and the new
integration variable $x=h\nu /kT$. Then Eq.(\ref{twentynine}) takes the form
\begin{equation}
\ln Z=-\frac 98N\left( \theta _D/T\right) -9N\frac 1{\left( \theta
_D/T\right) ^3}\int_0^{\theta _D\ \!/T}x^2\ln \left[ 1-\exp \left( -x\right)
\right] dx.  \label{thirtyone}
\end{equation}

\noindent Integration by parts of Eq.(\ref{thirtyone}) gives
\[
\fl
\ln Z=-\frac 98N\left( \theta _D/T\right) -3N\ln \left[ 1-\exp \left(
-\theta _D/T\right) \right] +3N\frac 1{\left( \theta _D/T\right)
^3}\int_0^{\theta _D\ \!/T}\frac{x^3dx}{e^x-1}.
\]

\noindent The last equation can be written in a more compact form if we
introduce the function
\[
D(z)=\frac 3{z^3}\int_0^z\frac{x^3dx}{e^x-1},
\]

\noindent which is known as the Debye function and $z=\theta _D/T$. Then the
logarithm of the partition function takes the form
\begin{equation}  \label{thirtytwo}
\ln Z=-\frac 98N\left( \theta _D/T\right) -3N\ln \left[ 1-\exp \left(
-\theta _D/T\right) \right] +ND\left( \theta _D/T\right) .
\end{equation}

From Eq.(\ref{thirtytwo}) and the thermodynamic identities
\[
\begin{array}{ll}
F=-kT\ln Z, & E=F-T(\partial F/\partial T)_{V,} \\
P=-(\partial F/\partial V)_T, & C_V=(\partial E/\partial T)_V
\end{array}
\]

\noindent we can derive expressions for the specific free and internal
energy, the pressure and the specific heat:
\[
\begin{array}{lll}
F_{vib}(T,V)\! & = & \!\frac 98Nk\theta _D+NkT\left\{ 3\ln \left[ 1-\exp
\left( -\theta _D/T\right) \right] -D(\theta _D/T)\right\} ,
\end{array}
\]
\[
\begin{array}{lll}
E_{vib}(T,V)\! & = & \!\frac 98Nk\theta _D+3NkTD(\theta _D/T),
\end{array}
\]
\[
\begin{array}{lll}
C_V\! & = & \!3Nk\left[ D(\theta _D/T)-(\theta _D/T)D{}^{\prime }(\theta
_D/T)\right] ,
\end{array}
\]

\noindent where (9/8){\em Nk}$\theta _D$ is the contribution of the quantum
zero oscillations in terms of the Debye model, and $D^{\prime }$ is the
derivative of {\em D} with respect to $\theta _D/${\em T}.

In order to derive {\em P}$_{vib}$ it is necessary to compute the volume
derivative of ln {\em Z}. The partition function does not depend explicitly
on volume but through the dependence of the energy levels on it. It is not
possible to obtain this relation from first principles. This imposes the use
of various approximations. In the case of a solid the volume dependence of
the energy levels is reduced to volume dependence of the lattice
frequencies. This relation cannot be obtained from first principles as well.
According to Gr\"uneisen
\begin{equation}  \label{thirtythree}
d\ln \nu _j/d\ln V=-\gamma \quad (j=1,2,...,3N),
\end{equation}

\noindent where $\nu _j${\em (V)} are lattice frequencies, and $\gamma $ is
a positive quantity, one and the same for all the 3{\em N} normal modes of
the crystal lattice. It is assumed that these frequencies, and hence, $%
\gamma $, do not depend on temperature, but on volume only. This is
sometimes referred to as the quasiharmonic approximation.

Equation (\ref{thirtythree}) is the statistical definition of the
Gr\"{u}neisen parameter. It holds for any frequency, hence it may be written
for the Debye frequency $\nu _D$%
\[
d\ln \nu _D/d\ln V=-\gamma
\]

\noindent and since $\nu _D$ and $\theta _D$ are proportional, it follows
\[
d\ln \theta _D/d\ln V=-\gamma \quad \ \mbox{\rm or\quad }\ d\ln \theta
_D/dV=-\gamma /V.
\]

\noindent Accordingly, we obtain for {\em P}$_{vib}$%
\[
P_{vib}=\frac 98\left[ \gamma \left( V\right) /V\right] Nk\theta _D+3\left[
\gamma \left( V\right) /V\right] NkTD(\theta _D/T)
\]

\noindent or
\[
P_{vib}=\left[ \gamma (V)/V\right] E_{vib},
\]

\noindent which in terms of the assumed approximation $\gamma /V=\gamma
_0/V_0=const.$ takes the form
\[
P_{vib}=(\gamma _0/V_0)E_{vib}.
\]

These results render it possible to write Eqs.(\ref{seven}) in an explicit
form:
\[
F(T,\varepsilon )=\sum_iE_{0i}\varepsilon ^i+\frac 98Nk\theta _D+NkT\left\{
3\ln \left[ 1-\exp \left( -\theta _D/T\right) \right] -D(\theta
_D/T)\right\} ,
\]
\[
E(T,\varepsilon )=\sum_iE_{0i}\varepsilon ^i+\frac 98Nk\theta
_D+3NkTD(\theta _D/T),
\]
\[
P(T,\varepsilon )=\sum_iP_{0i}\varepsilon ^i+\frac 98(\gamma _0/V_0)Nk\theta
_D+3(\gamma _0/V_0)NkTD(\theta _D/T),
\]

\noindent where the coefficients {\em P}$_{0i}$ are given by Eq.(\ref
{twentyeight}).

These are the complete, the caloric and the thermal EOS for solids. They are
defined over domains of the ({\em P},$\varepsilon $) and ({\em E},$%
\varepsilon $) planes formed by the set of experimental values of $%
\varepsilon $, {\em P} and {\em E}.

The integration constant {\em E}$_{00}$ can be easily obtained from the
caloric EOS. Let us write it in explicit form:
\[
E(T,\varepsilon )=E_{00}+E_{01}\varepsilon +E_{02}\varepsilon ^2+...+\frac
98Nk\theta _D+3NkTD(\theta _D/T).
\]

\noindent Since ambient conditions are assumed as a reference state, e.g.
{\em E}({\em T}=300{\em K},$\varepsilon $=0) = 0, we have
\[
E_{00}=-\frac 98Nk\theta _D-900NkD(\theta _D/300).
\]

Knowing the complete EOS, we are able in principle to derive all the
thermodynamic properties of the system under consideration by simple
differentiation.

To calculate the temperature on the shock Hugoniot we may use either the
caloric or the thermal EOS. Let us write the thermal EOS in the form {\em T
= f(T)} and replace {\em P}({\em T},$\varepsilon $) with {\em P}$_H$($%
\varepsilon $). We obtain
\[
T_H(\varepsilon )=(V_0/3\gamma _0)\left[ P_H(\varepsilon
)-\sum_iP_{0i}\varepsilon ^i-\frac 98(\gamma _0/V_0)Nk\theta _D\right]
/NkD(\theta _D/T_H).
\]

\noindent This equation may be solved either graphically or by the method of
successive approximations.

The simultaneous solution of the caloric and the thermal EOS in order to
eliminate {\em T} defines the incomplete EOS
\[
P(E,\varepsilon )=(\gamma _0/V_0)E+\sum_i\left\{ \left[ \left( i+1\right)
/V_0\right] E_{0i+1}-(\gamma _0/V_0)E_{0i}\right\} \varepsilon ^i.
\]

This is the constitutive relation necessary to supplement the system of
equations of fluid dynamics describing adiabatic compressible inviscid flow.

\section{Conclusions}

The assumptions that the free energy of a solid may be represented as a
superposition of terms, corresponding to the cold compressed state, the
quantum zero vibrations and the thermal vibrations of the crystal lattice,
that the ratio $\gamma (V)/V$ remains constant, and the Debye model for the
specific heat, along with the experimental shock Hugoniot allow the
derivation of its thermodynamic properties in explicit form. In the
beginning a differential equation for the zero Kelvin isotherm is derived
using the Mie-Gr\"uneisen equation and the shock Hugoniot as a reference
curve. An analytic solution to this equation is obtained in the form of
power series. The arbitrary constant is expressed in terms of the specimen's
parameters at ambient conditions. Based on this solution, the complete, the
caloric, the thermal and the incomplete EOS are obtained.

   The proposed EOS are valid in the range of pressures and temperatures
where the behaviour of the solid state is determined by the lattice thermal
vibrations. At very low and high temperatures the conduction electrons
contribution must be taken into account. At higher temperatures the
anharmonicity effects should be included as well.

   The complete EOS is derived mainly within the framework of thermodynamics
and it cannot predict phase transitions. Should a phase transition (or
melting) be registered experimentally, the high-pressure phase (the melt)
might be described using the technique, given in \cite{Zharkov}.

   The obtained incomplete EOS derives from a thermodynamically consistent
complete EOS and, therefore, it might serve as a reliable physical
constitutive relation, required to elucidate how the characteristics of the
material affect the propagation of shock waves. This is one of the aims of
the present work.

Also, an equation for the shock temperature along the Hugoniot curve
is derived.

Without the use of the algebraic programming system {\small REDUCE}
\cite{Hearn} many of these expressions could not have been obtained
easily.

In a forthcoming paper the proposed EOS will be applied to some solids
for which the necessary shock-compression data are available.

\section*{Acknowledgements}

  The author would like to express his thanks to Prof. N.Martinov,
D.Sci., Head of Condensed Matter Physics Department, Faculty of
Physics, Sofia University, for his careful reading of the manuscript and the
valuable suggestions, and also to  BALKAN AIRTOUR Company for the
computational and technical support.

\end{document}